\documentclass[preprint,showpacs,preprintnumbers,amsmath,amssymb]{revtex4}
\usepackage{amssymb}

\usepackage{graphicx}
\usepackage{dcolumn}
\usepackage{bm}

\begin{document}

\preprint{}

\title{Self-Induced Fractional Fourier Transform
 and Revivable Higher Order Spatial Solitons in Strongly Nonlocal Nonlinear Media}

\author{Daquan Lu}
 \author{Wei Hu}%
 \email{huwei@scnu.edu.cn}
 \author{Yajian zheng}
 \author{Yanbin liang}
 \author{Longgui Cao}
 \author{Sheng Lan}
 \author{Qi Guo}
\affiliation{Laboratory of Photonic Information Technology, South
China Normal University, Guangzhou 510631, China }

\date{\today}

\begin{abstract}

The fractional Fourier transform (FRFT) naturally exists in the
strongly nonlocal nonlinear (SNN) media  and the propagation of
optical beams in SNN media can be simply regarded as a self-induced
FRFT. Through FRFT technic the evolution of fields in SNN media can
be conveniently dealt with and an arbitrary square-integrable  input
field presents generally as a revivable higher order spatial soliton
which reconstructs  its profile periodically after every 4 times of
Fourier transforms. The self-induced FRFT would illuminate the
prospect of the SNN media in new applications such as continuously
tunable nonlinearity-induced FRFT devices.
\end{abstract}

\maketitle
Since Fourier suggested the usage of Fourier analysis to solve the
heat conduction problem in 1807, the Fourier transform (FT) has been
applied widely in many branches of science \cite{a0}. In the field
of optics, the FT is one of the most important and basic tools in
dealing with physical optics and optical information processing
\cite{a0}. In fact, the term Fourier optics is often used
synonymously with optical information process. The fractional
Fourier transform (FRFT), which is an extension  of the FT
\cite{namias}, has been introduced to optics since 1993 when
Mendlovic and Ozaktas find this operator can be optically performed
by the quadratic graded-index (GRIN) media \cite{a1}. In respect
that the FRFT can show the characteristics of the signal changing
continuously from the spatial domain to the spectrum domain, it
interests optics scientists and engineers and plays an important
role in many optics fields \cite{a2,a3,a4,a5,a6,a8,a9,a10,Finet94},
such as diffraction \cite{Finet94}, transmission \cite{a2,a3},
imaging \cite{a4}, information processing \cite{a6,a10}, etc..

On the other hand, since  the first observation of nonlinear optical
phenomena by Franken et al in 1961 \cite{franken61}, one year after
the invention of the laser, the nonlinear optics has been a rapidly
expanded active field and widely influenced other fields. A special
branch of the nonlinear optics, the field of optical solitons,  have
grown enormously in the past decades. In particular, the soliton in
strongly nonlocal nonlinear (SNN) media which supports
(2+1)-dimension solitons, have attracted extensive interest and been
widely investigated in the past few years
\cite{b2,b3,b4,b6,b62,b7,b8,b10,b12,b13,b14,ellipticons,Briedis05},
since Snyder and Mitchell  simplified the nonlocal nonlinear
Schr\"{o}dinger equation (NNLSE) to a linear model (called
Snyder-Mitchell mode (SMM) in our papers) in the SNN case and found
an exact Gaussian-shaped ``accessible soliton"\cite{b1}. And the
``accessible soliton"  has shown its interest in potential
applications such as photonic switching and logic gating \cite{b14}.

The Fourier optics and the nonlinear optics might seem to be
independent of each other in that the FT and the FRFT are  linear
transforms and well known as  technic provided to solve problems in
linear system. But we note that the SNN media, the propagation
equation in which can be mathematically simplified to the linear SMM
\cite{b1}, would provide an opportunity to intersect them with each
other.

In this letter, the FRFT induced by the input field itself through
the SNN effect is introduced. It is shown that the FRFT naturally
exists in the SNN media. The  order of the self-induced FRFT in the
SNN media can be steered  with the input power in addition to the
propagation distance $z$, quite different from the traditional
linear (intensity-independent) FRFT devices such as the lens series
and the quadratic GRIN media. Due to the self-induced FRFT,
propagation of  an arbitrary square-integrable input field presents
generally as a {\it revivable higher order spatial soliton} (RHOSS)
which is similar to the higher order (1+1)D temporal soliton in
fiber. The build-block-like solitons, build-block-like breathers,
and their interaction, can be regarded as special cases of the RHOSS
and can be conveniently investigated with the FRFT technic.


The propagation of beams in nonlocal nonlinear media can be
phenomenologically described  by the NNLSE: $ 2ikn_0{\partial_z
\Phi} + n_0 \Delta_\bot \Phi + 2k^2{\triangle n} \Phi=0,$ where $k$
represents the wave number in the media with the linear part of the
refractive index $n_0$ when the nonlinear perturbation of refractive
index $\triangle n $ equals zero,  $\triangle n = n_2 \int {R(\vec r
- \vec r_a)} |\Phi|^2 {\rm{d}}^2 \vec {r}_a$ ($n_2$ is the nonlinear
index coefficient and $R$ is the normalized symmetric real spatial
response function of the media). In the case of SNN media we need
only keep the first two terms of the expansion of $\triangle n$ and
the NNLSE is simplified to the SMM \cite{b1}
\begin{equation}
2ik{\partial_{z}  \Phi } +  \Delta _ \bot \Phi - k^2 \gamma^2 P_0
r^{ 2} \Phi = 0,\label{2}
\end{equation}
where   $\gamma$ is a material constant, $P_0 = \int |\Phi|^2
{\rm{d^2}}\vec r $ is the input power.

We assume $\Phi (\vec r,z) = a(\vec r)\exp ( - i\beta z)$ to seek
the stationary solutions of Eq. (\ref{2}). Substituting this
expression into Eq. (\ref{2}) gives
 $
 2\beta ka= k^2 \gamma^2 P_0
r^{ 2}a - \Delta _ \bot  a$,  the eigen solutions of which are
Hermite Gaussian soliton family
\begin{equation}
a^{(p)}=HG_{m,n}^{(p)}  = c^{(p)}H_{m }
(\frac{x}{w_c})H_{n}(\frac{y}{w_c})
  e^{- \frac{{r^{ 2} }}{{2w_c^2 }}}
   \ \ \ \ \ \ \ \ \ \ \ \ \ \ (m+n=p)
\label{hermite}
\end{equation}
in cartesian coordinate \cite{b8}, Laguerre Gaussian soliton family
\begin{equation}
a^{(p)}=\left\{ \begin{array}{l}
LG_{l,q}^{c(p)}=c^{(p)}(\frac{r}{w_c})^{l} L_q^{l}
(\frac{r}{w_c})\cos (l\theta )
  e^{- \frac{{r^{ 2} }}{{2w_c^2 }}}\\
  LG_{l,q}^{s(p)}=c^{(p)}(\frac{r}{w_c})^{l} L_q^{l} (\frac{r}{w_c})\sin(l\theta
  ) e^{- \frac{{r^{ 2} }}{{2w_c^2 }}}
\end{array}\right.
\ \ \ \ \ \ \ (2q+l=p) \label{lagurrel}
\end{equation}
 in circular cylindrical coordinate \cite{b8}, or Ince Gaussian
 soliton family
in elliptical coordinate \cite{b62,ellipticons}. In Eqs.
(\ref{hermite}) and (\ref{lagurrel}), $p=0,1,2...$ is the order of
the solution with which the soliton eigenvalue can be
straightforwardly obtained:
\begin{equation}
\beta^{(p)}=(p+1)\beta_0,\label{7}
\end{equation}
$c^{(p)}$ is the normalized coefficient  ensures $ \int |
a^{(p)}|^2{\rm{d^2}}\vec r=P_0$, $w_c = ({{k^2 \gamma^2 P_0
}})^{-1/4} $ is a generalized beam width, $H_{m }(x)$and
$L_q^{l}(r)$ respectively represents the Hermite and the Laguerre
polynomials,
$\beta_0=\sqrt
{ P_0 }\gamma=1/kw_c^2 $.

Mathematically,  the profile of the eigen soliton solution,
$a^{(p)}$, is simultaneously the eigen function of the FRFT
\cite{namias,a1,a9}
\begin{equation}
\hat{F_\alpha  }\{ g(\vec r_1)\}(\vec r_2)  = \frac{{\exp [ i(\alpha
-\frac{\pi}{2} )]}}{{2\pi w_c^2\sin \alpha }}\exp [\frac{{ir_2^{ 2}
}}{{2 w_c^2\tan \alpha }}]\nonumber
  \int
 \exp[\frac{ ir_1^{ 2} }{2w_c^2\tan \alpha }
 -\frac{i\vec r_1 \cdot \vec r_2}{w_c^2\sin \alpha }]
 {g}(\vec r_1){\rm{d^2}}\vec r_1,\label{frft}
\end{equation}
i.e., $a^{(p)}$  is the eigen solution of the equation $
\hat{F_\alpha  }\{ a^{(p)}(\vec r_1) \} = a^{(p)}(\vec r_2) e^{ -
ip\alpha }$ with the eigen value $e^{ - ip\alpha }$, where the order
of the FRFT is
\begin{equation}
\alpha  =\beta_0 z=  \sqrt {P_0 }\gamma z.\label{13}
\end{equation}
Therefore the field of the eigen soliton  at $z$ is connected with
that at the entrance plane through the FRFT: $ \Phi ^{(p)} (\vec
r_2,z) = \hat{F_\alpha  }\{ \Phi ^{(p)} (\vec r_1,0)\}  \times \exp
( - i\alpha ). $

An arbitrary square-integrable input field  can be expressed as a
linear superposition of an arbitrary one of the three families of
eigen soliton solutions: $ \Phi (\vec r_1,0) = \sum\nolimits_{p =
0}^\infty {c_p \Phi ^{(p)} (\vec r_1,0)}.\label{superposition} $
 According to the linearity of Eq.
(\ref{2}) and the FRFT, the propagation in the SNN media is
presented as the FRFT
 on the input field
\begin{equation}
\Phi  (\vec r_2,z) = \hat{F_\alpha  }\{ \Phi (\vec r_1,0)\} \times
\exp ( - i\alpha ).\label{12}
\end{equation}
In the special case $\alpha=\pi/2$, the propagated field at $z$ is
deduced to the well-known FT of the  input field, reads
\begin{equation}
\Phi  (\vec r_2,z) = \frac{-i}{2\pi w_c^2}\int
 { \Phi (\vec r_1,0)}
 \exp[ -\frac{i\vec r_1 \cdot \vec r_2}{w_c^2 }]
 {\rm{d^2}}\vec r_1.\label{ft}
\end{equation}
\begin{figure}[]
\includegraphics[width=9cm]{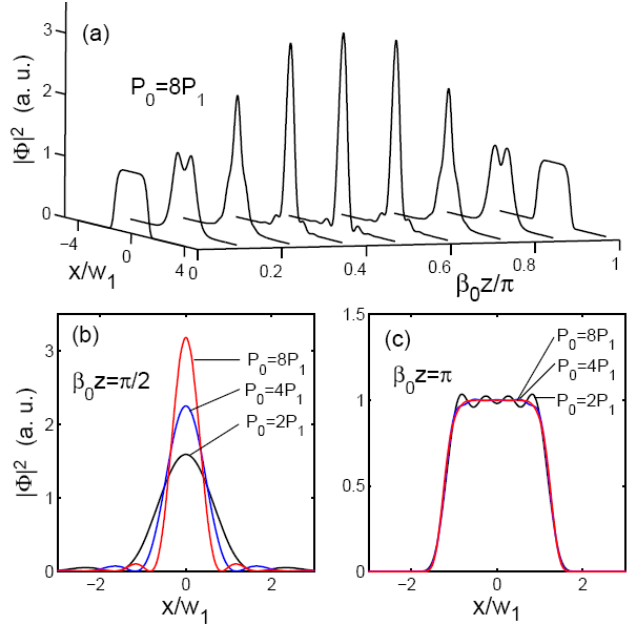}
\caption{\label{1d}
 (Color online)  Porpagation dynamics of the super Gaussian field
$\exp[-(x/\sqrt{2}w_1)^8]$ in SNN media with Gaussian response
function  $ R(r) = 1/(2\pi w_R^2 )\exp [ - r^{ 2} /(2w_R^2 )]$
\cite{b8,ellipticons,Briedis05}, based on  numerical simulation of
the NNLSE. (a) Evolution of the profile in propagation. (b), (c)
Intensity distribution 
at $\beta_0 z=\pi/2$ and $\pi$ respectively.  The nonlocality degree
$\Gamma=w_B/w_R=1/10$,  $w_B$ is the second-order moment width of
the beam at the entrance plane, $P_1$ is the critical power for the
eigen soliton  with generalized width $w_1$.
 }
\end{figure}
Equations (\ref{13})-(\ref{ft}) indicate that the SNN media
naturally performs the FRFT and FT on the input field. The physical
origination of this effect is as follows: When a  beam is input into
a SNN media,  it would induce a quadratic GRIN channel in the medium
through the nonlocality. The propagation in the channel then
performs the FRFT and FT, just as in the traditional quadratic GRIN
media\cite{a1}.  Because the grads of the refractive index
distribution can be steered with the power, the order of the
self-induced FRFT in the SNN meida can be steered  with the input
power in addition to the propagation distance $z$, quite different
from the traditional linear FRFT devices.


The simulation based on the NNLSE shows that:  the self-induced FRFT
becomes more and more distinct with the increase of the nonlocality
or the input power (Fig. \ref{1d}(c)). In fact, the increase of the
power is tantamount to the increase of the nonlocality, in that it
decreases $w_c$ and the average beam width in propagation.  On the
other hand, the smoother the shape of the input field is, the less
the nonlocality is required to support the FRFT, because it contains
less higher-order eigen soliton solutions in the superposition.
Generally, the profile of the eigen soliton  holds when
$\Gamma<1/10$. When the order  of a constituent eigen soliton in the
superposition is high enough so that $\Gamma >1/10$, the profile
would be distorted in propagation. Therefore the SNN media act as a
low-pass spatial frequency filter.

\begin{figure}[t]
\includegraphics[width=9cm]{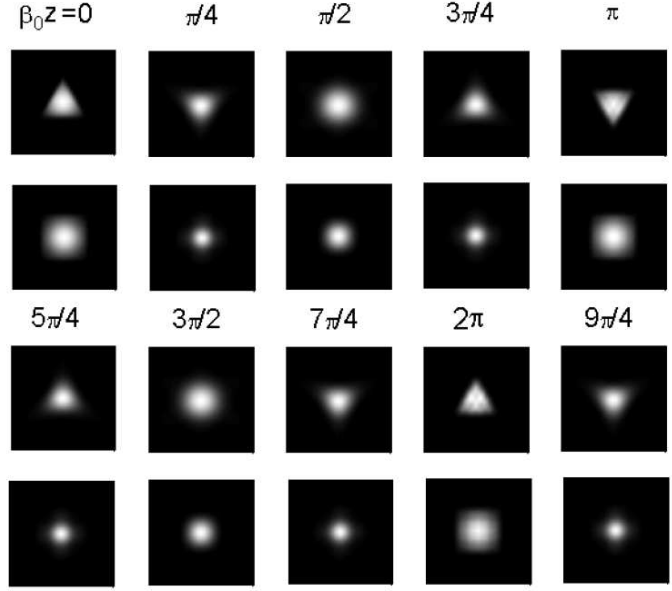}
\caption{\label{truncate} Propagation dynamics of Gaussian beams
($\exp[-(x/\sqrt{2}w_1)^2]$) truncated by  triangular (rows 1 and 3)
and square (rows 2 and 4) super Gaussian diaphragm in SNN media with
Gaussian response function based on   numerical simulation  of the
NNLSE.
$P_0=4P_1$. $\Gamma=1/20$.
 }
\end{figure}
Based on the self-induced FRFT, it is convenient to investigate the
propagation in SNN media from the angle of Fourier optics. We can
predict the field would present a periodical evolution with the
period $\Delta z = 2\pi /\sqrt{ P_0 }\gamma$ (corresponding to
$\Delta\alpha=2\pi $) (Figs. \ref{1d}-\ref{truncate}). As a result
of cascade FT, the pattern evolves to the inversion of the input
pattern at $z = (2n + 1)\pi /\sqrt { P_0 }\gamma$ and revive to the
input pattern at $z = 2n\pi /\sqrt { P_0 }\gamma$ (Fig.
\ref{truncate}) (we call those cross sections the revived planes or
imaging planes). At $z = (n + 1/2)\pi /\sqrt {
 P_0 }\gamma$ (we call these cross sections the Fourier planes),
the patterns are the FT spectrum of the input field or the
inversion. From Eq. (\ref{ft}) the spatial frequency $\vec k_r =
\vec r_2 /w_c^2$, thus at the Fourier planes the beam width $w(z)
\propto 1/\sqrt {P_0}$, which obeys the scaling rule of FT (Fig.
\ref{1d}(b)). Because of the periodic revivable evolution, which is
similar to the higher order temporal soliton in fiber, we call this
type of propagation the RHOSS (The RHOSS should be distinguished
from the traditionally mentioned ``higher order spatial soliton"
which refers to the stationarily propagated multipole spatial
soliton). There is an interesting difference between the higher
order temporal soliton in fiber and the RHOSS: the existence of the
higher order temporal soliton in fiber requires much more power than
that required for the fundamental soliton, whereas to support the
RHOSS, the power can be lower than the critical power of the
stationarily propagated fundamental  soliton.

\begin{figure}[b]
\includegraphics[width=9cm]{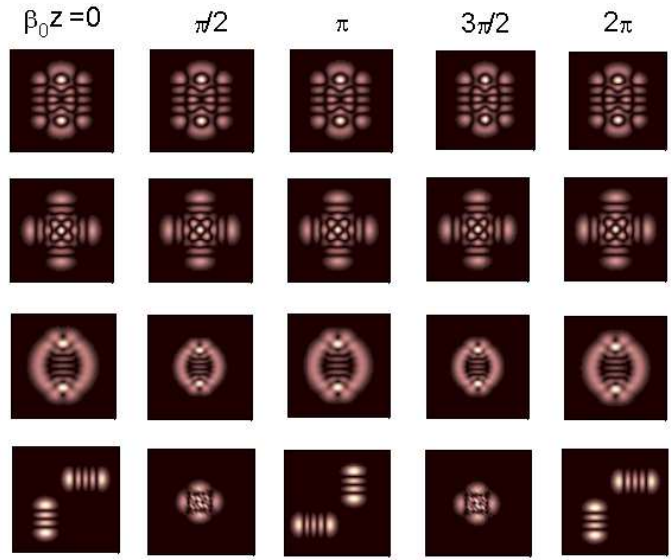}
\caption{\label{fig:buildblock} (Color online)
 Propagation dynamics  of
build-block-like soliton (rows 1-2), build-block-like breather (row
3), and multisoliton interaction (row 4) in SNN media with Gaussian
response function based on  numerical simulation of the NNLSE.
$\Gamma =1/10$, $P_0=2P_1$ for row 3 and $P_0=P_1$ for others. The
input fields are respectively $(HG^{(8)}_{4,4}-HG^{(8)}_{0,8}/10)$
(row 1), $(HG^{(8)}_{0,8}+HG^{(8)}_{8,0})$  (row 2),
$(LG^{(c,8)}_{8,0}+iLG^{(s,8)}_{8,0}-HG^{(8)}_{0,8}/125)$ (row 3)
and
$[HG^{(3)}_{0,3}(x+3w_1,y+3w_1)+HG^{(4)}_{4,0}(x-3w_1,y-3w_1)]$(row
4). The general width of all   constituent beams are  $w_1$ at the
entrance plane.}
\end{figure}

 According to properties of the FRFT, there are
three special cases of the RHOSS (Fig. \ref{fig:buildblock}):

\noindent {\it  1)  Build-block-like soliton.} \ As shown in Eqs.
(\ref{7}), during propagation, every degenerate eigen soliton  with
the order $p$ has the same propagation-induced phase $\beta^{(p)} z$
(or in other words, has the same FRFT eigen value $\exp
(-ip\alpha)$). Therefore, when i) the input field is the linear
superposition of the eigen soliton solutions with the same order
$p$, beam center, and generalized width $w_1$; and ii) the
 power is the critical power $P_1  = 1/(k^2 w_1^4 \gamma ^2
)$ which supports the eigen solitons with the generalized width
$w_1$, the field would propagate stationarily, i.e., the soliton
occurs (rows 1-2 in Fig. \ref{fig:buildblock}). Because the field of
this type can be freely composed of the eigen soliton solutions with
the same order, we call it build-block-like soliton.

\noindent {\it{2) Build-block-like breather.}} When all are the same
as the build-block-like soliton except the input power $P_0$ deviate
from the critical power $P_1$, the FRFT  keeps the shape of the
input field and periodically varies the width  with the period
$\Delta \alpha  = \pi$. Subsequently the evolution is presented as
breather with the period $\Delta z = \pi /\sqrt { P_0 }\gamma$ (row
3 in Fig. \ref{fig:buildblock}), and the change of the
build-block-like breather's width is the same as the prediction for
the Gaussian breather in Ref. \cite{b1}. At $z = (n + 1/2)\pi /\sqrt
{ P_0 }\gamma$, the FRFT is deduced to the FT and $w(z) \propto
1/\sqrt {P_0 }w_1$.

\noindent {\it 3) Multi soliton interaction.}
In this case, the traditional technic might  require much effort in
mathematical treatment because orbits of the interacting solitons
are influenced by each other. But by introducing the FRFT, the
evolution of the orbits is presented simply as a shift in the FRFT
reads  $\Phi {}(\vec r_2,z) =\hat{F_\alpha  } \{ \Phi {}(\vec r_1 -
\vec r_{0} ,0)\} \exp [ - i\alpha]\nonumber$,
where $\vec r_{0}$ represents the initial deviation of the beam
center of the interacting soliton   from the mass center. Under the
vertical incidence condition (row 4 in Fig. \ref{fig:buildblock}),
the solitons intersect each other at $z = (n + 1/2)\pi /\sqrt { P_0
}\gamma $, evolve to the inversion $\Phi {}_(\vec r_{0} - \vec r,0)$
at $z = (2n + 1)\pi /\sqrt { P_0 }\gamma $, and recur to the input
field $\Phi {}_(\vec r - \vec r_{0},0)$ at $z  = 2n\pi /\sqrt { P_0
}\gamma $.


\begin{figure}[t]
\includegraphics[width=9cm]{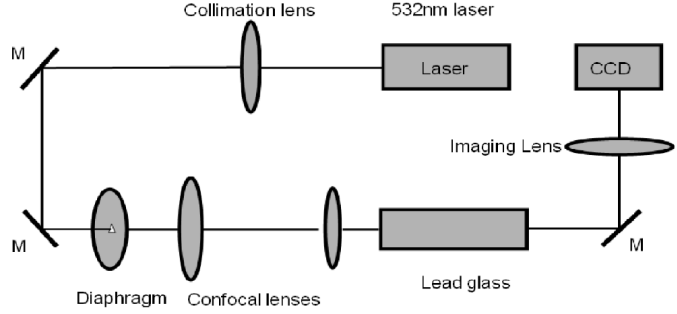}
\caption{The sketch of the experiment setup.\label{fig:setup} }
\end{figure}
\begin{figure}[t]
\includegraphics[width=9cm]{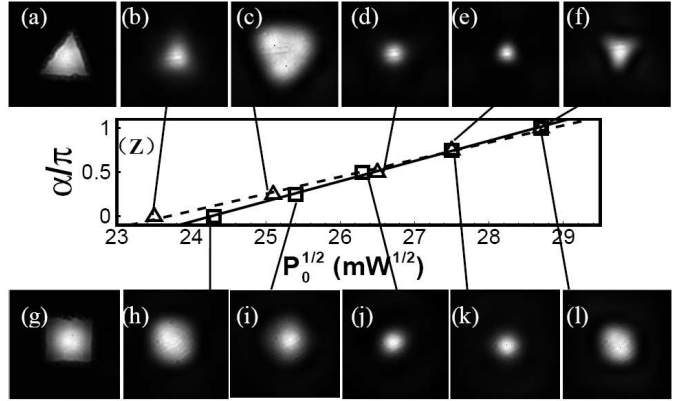}
\caption{\label{fig:experiment} Experimental results demonstrating
self-induced FRFT of Gaussian beams truncated by the triangular
((a)-(f)) and the square ((g)-(l)) diaphragm at different input
power. (a) and (g) are the input fields,  (b)-(f) and (h)-(l) are
the output patterns.
}
\end{figure}

To verify the prediction about the self-induced FRFT (or in other
words, the RHOSS), we carried out the experiment in a columned lead
glass. The experimental setup is illustrated in Fig.
\ref{fig:setup}. The beam from a Verdi laser is focused by the
collimation lens. A diaphragm is placed at the focus and the real
image is produced at the entrance plane of the lead glass by the
confocal lenses pair. When the input power is adjusted, the
intensity distribution at the entrance and exit plane are monitored
by imaging the beams onto a CCD camera. The lead glass is 59.8 mm in
length and 15.1 mm in diameter. The beam at the entrance of lead
glass is  85 $\mu$m $\times$ 80 $\mu$m in size (for square
diaphragm) or 106 $\mu$m in diameter of circumcircle (for triangle
diaphragm).

The experimental results are shown in Fig.
\ref{fig:experiment}(a)-(l): by changing the input power, the $w_c$
changes and the patterns similar to the FRFT spectrums with the
 orders $\alpha=0,$ $ \pi/4, $ $\pi/2, $ $3\pi
/4,$ and $ \pi$ in Fig. \ref{truncate}  are recorded. In case of the
triangle (square) diaphragm, the output pattern recur to the input
pattern when $P_0=551$ mW (Fig. \ref{fig:experiment}(b)) ($P_0=590$
mW (Fig. \ref{fig:experiment}(h))) and evolves to the inversion when
$P_0=823$ mW (Fig. \ref{fig:experiment}(f)) ($P_0=810$ mW (Fig.
\ref{fig:experiment}(l))). Because the higher-frequencies are
filtrated in propagation, the output patterns  are  smoother than
the input ones. In Fig. \ref{fig:experiment}(z), the variation of
the FRFT order $\alpha$  with the square of the  input power, i.e.
$\sqrt{P_0}$, is illustrated. The linear fit shows that the FRFT
order $\alpha$ is directly proportional to $\sqrt{P_0}$, as
predicted in Eq. (\ref{13}).

In summary, the self-induced FRFT in SNN media made the nonlinear
optics and the Fourier optics intersect with each other.The
introducing of the FRFT technic would release one from complicated
mathematical calculation in propagation problems such as soliton
solutions in SNN media.  The RHOSS, including the build-block-like
solitons, build-block-like breathers, and their interaction, would
greatly enrich the nonlocal soliton family.  The fact that the order
of the self-induced FRFT in SNN media is related not only to the
propagation distance but also to the input power, quite different
from that in the traditional linear devices,
would illuminate the prospect of new applications of the SNN media
such as developing power-controlled continuously tunable FRFT
devices.

This research was supported by the National Natural Science
Foundation of China (Nos. 10674050 and 10804033), the Program for
Innovative Research Team of the Higher Education in Guangdong (No.
06CXTD005), and Specialized Research Fund for the Doctoral Program
of Higher Education (No. 20060574006).

\end{document}